\documentstyle[12pt,epsf,epsfig]{article}
\def\beq{\begin{equation}}  
\def\eeq{\end{equation}}
\def\lsim{\mathrel{\rlap{\lower3pt\hbox{\hskip0pt$\sim$}}
    \raise1pt\hbox{$<$}}}         
\def\gsim{\mathrel{\rlap{\lower4pt\hbox{\hskip1pt$\sim$}}
    \raise1pt\hbox{$>$}}}         
\def\simlt{\mathrel{\raise.3ex\hbox{$<$\kern-.75em\lower1ex\hbox{$\sim$}}}}
\def\simgt{\mathrel{\raise.3ex\hbox{$>$\kern-.75em\lower1ex\hbox{$\sim$}}}}

\begin{document}
\begin{titlepage}

\begin{flushright}
hep-ph/0209003 \\
TPI--MINN--02/39\\
UMN--TH--2110/02 \\
August 2002
\end{flushright}
\begin{center}
\baselineskip25pt

\vspace{1cm}

{\Large\bf Preserving the Lepton Asymmetry \\in the Brane World}

\vspace{1cm}

{\sc
Durmu{\c s} A. Demir$^1$, Tony Gherghetta$^{2}$, Keith A. Olive$^{1,2}$}
\vspace{0.3cm}

$^1${\it Theoretical Physics Institute,
University of Minnesota, Minneapolis, MN 55455}

$^2${\it School of Physics and Astronomy, University of Minnesota,
Minneapolis MN 55455}
\end{center}
\vspace{1cm}
\begin{abstract}
In models where the Standard Model spectrum is localized
on a brane embedded in a higher dimensional spacetime,
we discuss the lepton number violation induced by the emission
of right-handed neutrinos from the brane. We show that 
the presence of the right-handed neutrinos
in the bulk may lead to rapid lepton number violating processes
which above the electroweak scale would wash away any
prior lepton or baryon asymmetry. We derive constraints on the Yukawa
couplings of these states in order to preserve the lepton asymmetry. We
show that this has a natural interpretation in the brane world.

\end{abstract}

\end{titlepage}

\section{Introduction}
The generation and survival of a baryon asymmetry is a requirement of 
any realistic cosmology and hence enables one to put
strong constraints on phenomenological models beyond the standard model.
While all three ingredients~\cite{sak} necessary for the generation of
a baryon asymmetry are contained within the standard model, it is well
established that electroweak baryogenesis can not provide an asymmetry of
sufficient magnitude \cite{noew}.   Even in the context of the 
supersymmetric standard model, baryon number generation is difficult
\cite{ewsusy} and requires a particular supersymmetric mass spectrum.
Therefore, the observed baryon asymmetry of the Universe (BAU) is likely
to arise from short-distance physics above the ${\rm TeV}$ scale. 
While sphaleron effects \cite{krs} may not be primarily responsible for
the generation of the BAU, they certainly do modify any pre-existing
asymmetry.  For example, a baryon asymmetry with $B - L = 0$, will
be washed away by sphaleron interactions \cite{am}. On the other hand, any
primordial $B-L$ asymmetry will be reshuffled by the rapid
anomalous ($B+L$)-violating  sphaleron transitions above the electroweak
phase transition temperature,
$T_c$. Thus a pure lepton asymmetry can be converted into a baryon
asymmetry \cite{fy}. However, any baryon/lepton number violating
interaction (other than purely $B+L$ violating interactions) in
equilibrium above $T_c$, will once again lead to the total erasure of both
$B$ and $L$. Thus, in order for successful leptogenesis to occur
\cite{fy}, the right-handed Majorana neutrino, whose out-of-equilibrium
decays generate the requisite
$B-L$ excess, must be massive enough to suppress lepton number violating
interactions above the weak scale  in order to preserve the asymmetry
against erasure \cite{fy2}.   

Going beyond the standard model, however, introduces the possibility of a
hierarchy of mass scales rendering the weak scale unstable to radiative
corrections \cite{hierarchy0}.  In addition to attempts at solving the
hierarchy problem \cite{gia1,rs1,hierarchy}, extended objects arising
in certain string vacua suggest that our universe is a (3+1)-dimensional
brane embedded in a higher dimensional spacetime. While all fields charged 
under the gauge group are generally supposed to be confined to a brane due to
the conservation of gauge flux, gauge singlets can in principle 
propagate freely in the bulk with far-reaching  consequences
\cite{bulkneutrino,giga}. In  the case of the graviton, one can lower
the fundamental scale of gravity down  to the  ${\rm TeV}$ scale thereby
providing a solution to the hierarchy problem \cite{gia1}. In the case of
right-handed neutrinos, one can naturally derive small Dirac masses for
neutrinos \cite{bulkneutrino,grossman,paul,other}.

Gauge singlets can be produced through the scatterings of 
brane-localized particles.  Once they are produced,
they propagate into the bulk with almost no probability of further 
interacting with  brane fields since the 
brane occupies only a tiny volume of the total space. 
These singlet-emission processes, therefore, show up as missing energy
signals in  colliders. 
There is, however, an important difference between the emission 
of gravitons and right-handed neutrinos
in that the latter may appear not only as missing
energy but also as a source of lepton number violation on the brane. 
Given our discussion above, it is clear that right-handed neutrino
emission into the bulk can have important consequences regarding the BAU.

In this letter, we will show that brane-world models 
with bulk right handed neutrinos can lead to a devastatingly small BAU.
We further derive conditions on the neutrino sector in order
to protect the lepton and baryon asymmetries.
We apply these results to a number of specific brane-world scenarios.
We show that these conditions are a natural consequence of the brane
world.

\section{Lepton Number Violation in the Brane World}
We begin by discussing the phenomenon of $L$ violation
and its implications for the BAU for a generic configuration
where a three brane with coordinates $x_{\mu}$ ($\mu=0,\cdots,3$)
is embedded in a higher dimensional space with extra dimensions
$y_{a}$ ($a=1,\cdots,\delta$). We introduce bulk fermions
$\Psi^{\ell}(x,\vec{y})$ ($\ell=1,2,\cdots,n_s$) which possess lepton number
with the decomposition
\begin{eqnarray}
\label{psi}
\Psi^{\ell}(x,\vec{y})=\left(\begin{array}{c} \psi^{\ell}_L(x,\vec{y})\\ 
\psi^{\ell}_R(x,\vec{y})\end{array}\right)~,
\end{eqnarray} 
where $n_s$ is the number of fermion species.
The component fields $\Psi_{L,R}^{\ell}$ can be expanded in a 
complete set of functions  $f^{\ell\ \vec{n}}_{L,R}(\vec{y})$ with 
respective coefficients $\psi_{L,R}^{\ell\ \vec{n}}(x)$, where $\vec{n}$ is a set 
of integers; one for each extra dimension.  Dirac neutrino masses 
arise from the coupling of $\psi_{R}^{\ell\ \vec{n}}$ to the composite SM
singlet 
$\overline{\bf L}^{\alpha} {\bf H}^{c}$~\cite{bulkneutrino}
\begin{eqnarray}
\label{lag}
S=\sum_{\vec{n}}\ \int d^{4}x\ h^{\alpha \ell}_{\vec{n}}\ 
\overline{\bf L}^{\alpha}(x)\ {\bf H}^{c}(x)\ \psi_{R}^{\ell\ \vec{n}}(x)~,
\end{eqnarray}
where ${\bf L}^{\alpha}={\bf L}^{e,\mu,\tau}$ are the lepton doublets, 
${\bf H}$ is the Higgs field, and 
\begin{eqnarray}
\label{yukawa}
h^{\alpha \ell}_{\vec{n}}\equiv \lambda^{\alpha 
\ell} f^{\ell\ \vec{n}}_{R}(\vec{y}_0)~,
\end{eqnarray}
where $\vec{y}_0$ is the position of the brane, and $\lambda$ is a
$3\times n_s$ Yukawa matrix.  From the experimental perspective, the
solar  and atmospheric  neutrino anomalies can, respectively,  be
explained by  $\nu_{e}\leftrightarrow \nu_{\mu}$ \cite{sno} and 
$\nu_{\mu}\leftrightarrow \nu_{\tau}$ \cite{atm}  oscillations with small 
contributions from the sterile states $\psi_{L,R}^{\ell\ \vec{n}}(x)$. 
It is thus 
reasonable to assume that the couplings to the zero modes alone already
provide  a good fit to experiment with appropriate textures \cite{paul}.
That is,  the unitary matrix $U$ which performs the diagonalization $U
\left(h_{\vec{0}}  h^{\dagger}_{\vec{0}}\right) U^{\dagger}
=\mbox{diag.}\left({h^{1}_{\vec{0}}}^2,  {h^{2}_{\vec{0}}}^2,
{h^{3}_{\vec{0}}}^2\right)$ explains the neutrino  oscillation data to a
good approximation. Neutrino masses would then be given by
$m_{\alpha}={h^{\alpha}_{\vec{0}}}  \langle {\bf H}\rangle$ if there are
no mixings with the higher KK modes.  However, each  left-handed neutrino
$\nu_{\alpha}$ (the neutral component of ${\bf  L}^{\alpha}$) mixes
with the higher KK states via an angle $\theta_{\alpha}$  with 
\begin{eqnarray}
\label{angle}
\tan^{2}\theta_{\alpha}=\sum_{\vec{n}\neq \vec{0}} 
h_{\vec{n}}^{\alpha \ell} \frac{1}{m^2_{\ell\ \vec{n}}} {h^{\dagger}}^{\ell \alpha}_{\vec{n}}
\langle {\bf H}\rangle^{2}~,
\end{eqnarray}
where $m_{\ell\ \vec{n}}$ is the KK mass of $\psi_{L,R}^{\ell\
\vec{n}}(x)$. This angle must be small, $\tan\theta_{\alpha}\simlt 0.1$,
according to the experimental  constraints from reactor, accelerator,
solar and atmospheric neutrino data 
\cite{paul}. The solar and atmospheric anomalies, respectively, imply 
$\delta m_{sol}^{2}=m_2^2 - m_1^2 = 3.7\times 10^{-5}\ {\rm eV}^{2}$
and $\delta m_{atm}^{2}=m_3^2 - m_2^2 = 3.0\times 10^{-3}\ {\rm eV}^{2}$ 
assuming large mixings in both cases.  We note that although we  have
generated Dirac masses in this model, the cosmological constraint on
additional neutrino states \cite{oss} is not violated since the
right-handed states are necessarily much more weakly coupled than their
left-handed counterparts.  In what follows,  all  discussions will be
based on the effective action (\ref{lag}).

For now, we assume that a primordial $B-L$ asymmetry has been generated at
some high temperature  $T_{B-L}\gg T_c$. Below
$T_{B-L}$, brane matter and sphalerons are in thermal and chemical
equilibrium. 
By assigning each particle species a chemical potential, and 
using gauge and Higgs
interactions as conditions on these potentials 
(with generation indices suppressed), one obtains
\begin{eqnarray}
\mu_{d_L} - \mu_{u_L} = 0~; \qquad \mu_{l_L} - \mu_\nu =  0~; \qquad
\mu_{u_R} - \mu_{u_L} = -\mu_{d_R} + \mu_{d_L} = -\mu_{l_R}+\mu_{l_L} = \mu_H~,
\label{chems}
\end{eqnarray}
where the constraint on the weak isospin charge, 
$Q_3 \propto \mu_W = 0$ has been employed.
{}From (\ref{chems}), one can write 
down a simple set of equations for the baryon and 
lepton numbers and electric charge which reduce to:
\begin{eqnarray}
B& \propto & 4 N \mu_{u_L}~, \nonumber \\
L& \propto & 3 \mu - N \mu_H~, \label{mus} \\
Q& \propto & 2N \mu_{u_L} -2\mu + (4N+2) \mu_H~, \nonumber
\end{eqnarray}
where $\mu = \sum \mu_{\nu_{\alpha}}$ and $N$ is the number of generations.
Equilibrium sphaleron transitions further restrict
\beq
3 N \mu_{u_L}+ \mu=0~.
\label{shal}
\eeq
In the absence of any other
$B-L$ violating interactions (in equilibrium), these 
conditions  ultimately give
\beq
B = {28 \over 79} \left( B - L \right) ~,
\label{2879}
\eeq
for $N=3$ generations. Thus, in the absence of a primordial $B-L$ asymmetry, 
the baryon number is erased by equilibrium processes \cite{equilibrium}.
Note that barring new interactions (in an extended model)
the quantities ${1 \over 3}B - L_e$, ${1 \over 3}B - L_\mu$,
and ${1 \over 3}B - L_\tau$ remain conserved.

\begin{figure}[h]
\hspace*{-0.1in}
\begin{minipage}{3.0in}
\begin{center}
\hspace*{1.3in}
\epsfig{file=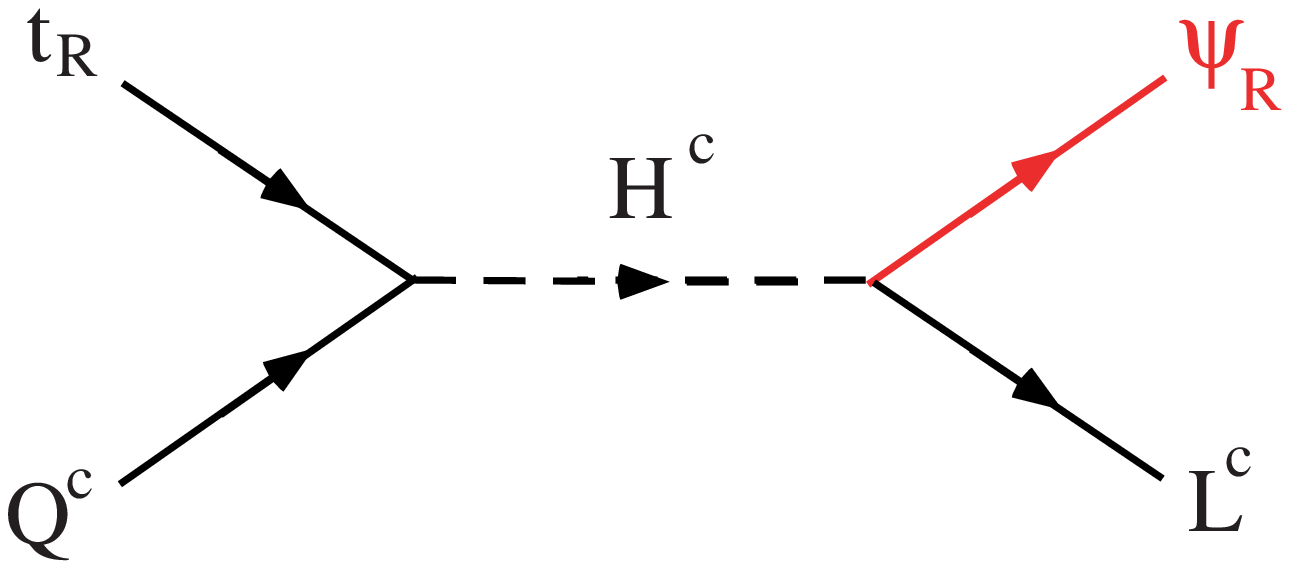,height=1.7in}
\end{center}
\end{minipage}
\vskip .2in
\caption{\label{fig1}
{The scattering of third generation quarks into $\psi_R {\bf L}^c$. The
right-handed neutrino,  with non-zero momentum along the extra dimension,
induces lepton number nonconservation on the brane.}}
\end{figure}

It is straightforward to see that the emission of right-handed neutrinos
from the brane into the bulk is a source of lepton number violation
which leads to the erasure of any $B-L$ asymmetry. At temperatures
above $T_c$, the dominant lepton number violating interactions are
the scattering of top quarks into right-handed neutrinos and
a lepton doublet, $Q^{c} t_{R} \rightarrow L^{c} \psi_{R}^{\vec{n}}$,
$Q^{c} L \rightarrow t_R^{c} \psi_{R}^{\vec{n}}$, and
$t_{R} L \rightarrow Q \psi_{R}^{\vec{n}}$, as depicted in Fig. 1.
At high temperatures, the production rates for a given neutrino
flavor and its SU(2) charged lepton partner are identical as the mixing
angle (\ref{angle}) vanishes. Therefore, the  thermal and  chemical
balance imposed by  fast electroweak interactions  between neutrinos
and their SU(2) partners still holds.  If the rates of these processes
are faster than the expansion rate of the Universe, while the reverse
process (being suppressed by the number of kinematically accessible KK
states) is not, they lead to an additional condition on the chemical
potentials, namely $\mu = 0$. In this case, one can easily see that all
of the chemical potentials are forced to zero as is the BAU.

We next consider this process in more detail. Given the collision processes
in Fig. 1 then the difference between the number densities of $\nu_{L}$
and $\nu_{L}^{c}$ of a given generation evolves as
\begin{eqnarray}
\label{dense}
&&\left(\partial_t + 3 H\right) (n_{\nu} - n_{\nu^c})= - \Gamma_F +
\Gamma_R +
\frac{1}{6}
\left(\partial_t - 3 H\right) (n_{b} - n_{b^c})~,
\end{eqnarray}
where the last term comes from the sphaleron contribution. 
The forward process is given by
\begin{eqnarray}
\Gamma_F & =  & \sum_{\vec n} \int \frac{d^3 
\vec{p}_{Q}}{2 (2\pi)^{3} |\vec{p}_{Q}|} \frac{d^3 \vec{p}_{R}}{2 (2\pi)^{3} 
|\vec{p}_{R}|} \frac{d^3 \vec{p}_{L}}{2 (2\pi)^{3} |\vec{p}_{L}|} 
\frac{d^3 \vec{p}_{\vec{n}}}{2 (2\pi)^{3} E_{\vec{n}}}\nonumber\\
&&\left\{ f_{Q^{c}} f_{t_{R}} (1-f_{L^c}) |{\cal{A}}(Q^{c} t_{R} \rightarrow 
L^{c} \psi_{R}^{\vec{n}})|^{2} + f_{Q^c} f_{L} (1-f_{t_{R}^c}) 
|{\cal{A}}(Q^{c} L \rightarrow t_R^{c} \psi_{R}^{\vec{n}})|^{2}\right.
\nonumber\\&&\left.+ f_{t_R} f_L (1-f_Q) |{\cal{A}}(t_R L \rightarrow Q 
\psi_{R}^{\vec{n}})|^{2}
\right\}+ {c.c.}
\end{eqnarray}
Here, we
follow the procedure and notation described in \cite{cko}. In
(\ref{dense}),
$H$ is the Hubble  expansion parameter and $f^{-1}=e^{(E-\mu)/T}+1$ is
the phase space distribution  for fermions. For the latter, $f\approx 
(1+\mu/T)/(e^{E/T}+1)$ is an excellent approximation since  the chemical
potential consistent with the BAU is small, $|\mu|\sim 10^{-10} T$. 
The reverse process, $\Gamma_R$ is phase space suppressed relative
to the forward process. The suppression is related to the relative
thickness of the brane, $\Delta$, and the size of the compact bulk
space, $R$, namely $\Gamma_R \simeq \max((\frac{\Delta}{R})^\delta, 
\frac{1}{(RT)^\delta})\Gamma_F$. 
Since the number of kinematically accessible KK states is given by 
$N_{KK} = (RT)^\delta$,
and we consider $T \lsim \Delta^{-1}$, we see that the suppression is given
simply by $\Gamma_R \simeq (1/N_{KK}) \Gamma_F$. In a standard 4D
picture with a single right-handed state, we clearly have $\Gamma_R
= \Gamma_F$ and since $\nu_R$ can be assigned a lepton number, there is
no lepton number violation in the theory. However, if $\Gamma_R \ne
 \Gamma_F$, there is an effective violation of lepton number if the
forward rate is fast compared with the expansion rate, $H$, while the
reverse rate is not.

In evaluating  (\ref{dense}), we approximate $f$ by the Maxwell-Boltzmann 
form,  and  take $1-f\approx 1$. The lepton asymmetry in $\nu_{L}$ is
defined by 
\beq 
L_{\nu}= {n_{\nu} - n_{\nu^c} \over s} = {15 \mu
\over 4 \pi^2 g_* T} + {\cal{O}}\left(\frac{\mu}{T}\right)^3~,
\eeq
where the entropy density, $s\approx \frac{2\pi^2}{45} g_{\ast} T^3$,
obeys $d s/dt = - 3 s H$ in an
adiabatically  expanding universe and $g_{\ast}$ is the number of
relativistic degrees of freedom;  equal to about $10^{2}$ for the standard
model spectrum. Then  direct evaluation of (\ref{dense}) gives
\begin{eqnarray}
\label{dense2}
\frac{d}{dt} (B-L) = \sum_{\vec{n}} \frac{3 h_t^2}{4(2\pi)^{5}}\ \frac{T^3}{s}\ \sum_{\alpha} 
\left(h_{\vec{n}}\  F\left(\frac{m_{\vec{n}}}{T}\right)\ h^{\dagger}_{\vec{n}}\right)_{\alpha \alpha} \left(\mu_H +\mu_{\nu_\alpha} 
\right)~,
\end{eqnarray}
where the sum over $\vec{n}$ extends up to the heaviest kinematically
accessible  KK mode; $m_{\vec{n}}^{max} \sim 2 \pi T$. Here the function
$F(x)$  varies slowly with $x$; it is $\approx 2$ for $x\ll 1$ and
$\approx 1$  for $x\approx 1$.

We next discuss the implications of (\ref{dense2}) which depend on the
scheme used to generate neutrino masses and oscillations. We distinguish 
two broad classes: ($i$) a hierarchy in the overlap of the wavefunctions, 
$f^{\ell\ \vec{0}}_{R}(\vec{y}_0)$, or ($ii$) 
a hierarchy in the Yukawa couplings, $\lambda^{\alpha \ell}$.

\subsection{Hierarchy from wavefunction overlap}
Here, we assume that the hierarchy among the
neutrino masses as well as the requisite textures for neutrino 
oscillations are both generated by the zero mode wave functions $f^{\ell\ 
\vec{0}}_{R}(\vec{y}_0)$. Namely, we take
the entries of the effective Yukawa coupling matrix $\lambda^{\alpha \ell}$ to
be of order one. This implies that $(h_{\vec{n}}
h^{\dagger}_{\vec{n}})_{1 1} 
\approx (h_{\vec{n}} h^{\dagger}_{\vec{n}})_{2 2} \approx (h_{\vec{n}} 
h^{\dagger}_{\vec{n}})_{3 3} \equiv h_{\vec{n}}^{2}$. 
Consequently (\ref{dense2}) reduces to 
\begin{eqnarray}
\label{dense3}
\frac{d}{dt} (B-L) = - \Gamma(T) (B - L)~,
\end{eqnarray}
with  
\begin{eqnarray}
\Gamma(T) = \sum_{\vec{n}} \frac{ h_t^2 h_{\vec{n}}^{2}}{ 20 (2\pi)^{3}}\
\frac{10 N +3}{22 N + 13}\ \frac{g_{\ast} T^4}{s}\ F~,
\label{gt}
\end{eqnarray}
where $F$ as noted above is {\cal O}(1), and we have expressed 
$N \mu_H + \mu$ in terms of $B-L$ using Eqs. (\ref{mus}) and (\ref{shal}). 
This equation implies that the emission of  right-handed neutrinos from the
brane drives the system towards
$\mu=-N \mu_H$, so long as the reverse process is sufficiently suppressed.
Thus if the rate of
$L$ violation is also fast compared to the expansion rate of the Universe
(at the same time sphaleron processes are in equilibrium), this
constraint, when combined with  other equilibrium conditions, drives {\em
all} asymmetries in the system to zero.  Indeed, the equilibrium relation
$\mu=-N \mu_H$ is the same condition which arises in the presence of
heavy right-handed neutrinos in four dimensions, and avoiding it places a
bound on the combination $h^2/M$ \cite{fy2,equilibrium,cdo}, where $h$ is
the right-handed Yukawa coupling and $M$ is the Majorana mass of the
right-handed neutrino.

A more quantitative determination of the damping of $B-L$ via (\ref{dense3})
as the temperature  falls to $T\sim T_c$ requires some
knowledge of the cosmological evolution at higher temperatures. 
We assume that there is a critical temperature
$T_{brane} \ge T_{B-L}$, below which the standard
FRW evolution holds. $T_{B-L}$ is the temperature at which a
$B-L$ asymmetry is produced and must be greater than $T_c$ (unless a
mechanism for producing a net baryon asymmetry below $T_c$ is provided).  
Thus our initial conditions are such that at $T=T_{brane}$, the bulk is 
empty, and the brane contains hot SM matter with an energy density
\begin{eqnarray}
\label{density}
\rho=\frac{\pi^2}{30}\ g_{\ast}\ T^{4}~,
\end{eqnarray}
where the expansion rate of the Universe is simply
$H^{2} = {\rho}/{3 M_{Pl}^{2}}$. 
The emptiness of the bulk is a reasonable assumption if $T_{brane}$ is the
reheat temperature, and the inflaton is a brane-localized field (See
$e.g.$ \cite{empty}). The emission of gravitons and right-handed
neutrinos depletes the energy  density on the brane beyond the dilution
due to expansion.  However, this  energy exchange between the brane and
the bulk does not affect $H$ which depends only on the sum of the brane
and bulk energy densities (See, $e.g.$ \cite{kiritsis}).  

\begin{figure}[h]
\hspace*{-0.1in}
\begin{minipage}{3.0in}
\begin{center}
\hspace*{1.3in}
\epsfig{file=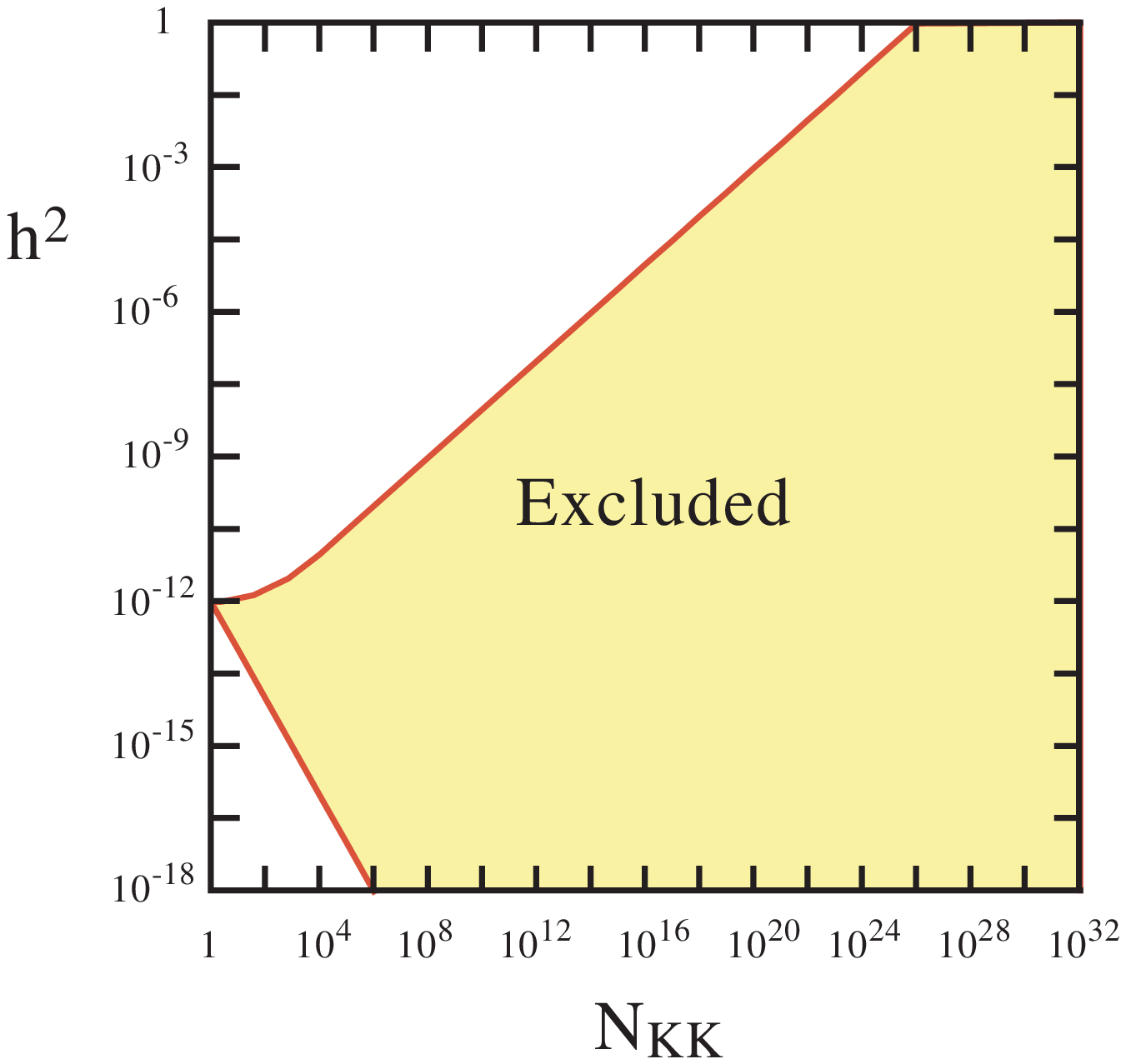,height=4in}
\end{center}
\end{minipage}
\vskip .2in
\caption{\label{fig2}
{The excluded region (shaded) in the $h^2-N_{KK}$ plane. Above the shaded
region inverse scatterings are rapid enough so that there is no
effective violation of lepton number.  Below the shaded region, even the
forward process is too slow and right-handed states are never produced.}}
\end{figure}

We next must determine if either the forward or backward rates are fast
compared with the expansion rate, $H$. Using Eq. (\ref{gt}), we see that
\beq
\frac{\Gamma_F}{H} \simeq 10^{-3} h^2 \frac{N_{KK}}{\sqrt{g_*}}
\frac{M_{Pl}}{T}~,
\eeq
assuming that the $h_n \sim h$ are all constant for the purposes of
making a numerical estimate. At $T_c$, $\Gamma_F/H \simeq 10^{13} h^2
N_{KK}/\sqrt{g_*}$. Thus for $h^2 N_{KK} \gsim 10^{-13}\sqrt{g_*}$, 
this rate will be rapid enough to produce $\nu_R$'s in the bulk. The 
reverse rate compared with the expansion rate is then simply 
$\Gamma_R/H \simeq 10^{13} h^2/\sqrt{g_*}$, so that for 
\beq
h^2 \gsim 10^{-12}\sqrt{1+\frac{N_{KK}}{100}}~,
\label{rapidr}
\eeq
this rate is also rapid and no lepton number violation occurs.  This is
our first main result.
When this bound is violated and at the same time $h^2 N_{KK} \gsim
10^{-12}$, then indeed the lepton and baryon number of the Universe is
erased.  Our second constraint is therefore 
\beq
h^2 \lsim 10^{-12}/N_{KK}~.
\label{rapidf}
\eeq
Preservation of the baryon asymmetry requires that either (\ref{rapidr})
or (\ref{rapidf}) is satisfied. These results are summarized in
Fig. \ref{fig2}. The constraint due to insuring that the reverse process
is rapid compared to $H$ and given by Eq. (\ref{rapidr}) defines the
upper boundrary of the excluded region. We note that had we included the
decays of the right-handed states, this exclusion bound would be slightly
weaker. At very large $N_{KK}$, it is softned by a factor of approximately
$T_{B-L}/T_c$. The lower boundary is derived by insuring that the
forward process is out of equilibrium (Eq. (\ref{rapidf})).

To see that the baryon asymmetry is effectively erased when both of
the above conditions are violated, we use
$H=1/2 t$,  appropriate for radiation domination, and relate the
temperature of the brane matter to time as
\begin{eqnarray}
\label{time}
t = \sqrt{45 \over 2 \pi^2 {g_{\ast}}}{M_{Pl} \over T^2} \; .
\end{eqnarray}
Using this expression and assuming entropy conservation,  
$B-L$ can be integrated from (\ref{dense3}) to obtain
\begin{eqnarray}
\label{result}
\frac{(B-L)[ T_c]}{(B-L)[T_{B-L}]}=e^{- \gamma_0
\frac{M_{Pl}}{T_c}\left(1-\frac{T_c}{T_{B-L}}\right)}~,
\end{eqnarray}
which gives us $B-L$ at $T=T_c$ relative to the primordial value of $B-L$
at $T=T_{B-L}$. The constant $\gamma_0$ in the exponent (\ref{result})
is given by 
\begin{eqnarray}
\gamma_0= \sum_{\vec{n}} \frac{27 h_t^2 h_{\vec{n}}^{2}}{(2\pi)^{6}}
~\sqrt{\frac{10}{g_{\ast}}}~
\frac{10 N + 3}{22 N +13} \sim 10^{-3} h^2 \frac{N_{KK}}{\sqrt{g_*}}~,
\end{eqnarray}
where each emitted KK mode contributes an amount $\sim 10^{-4} 
h_{\vec{n}}^{2}$ to the sum, so that the exponent in (\ref{result}) 
has the numerical value 
$\sim \sum_{\vec{n}} h_{\vec{n}}^{2}\ 10^{12}\ (1- T_c/T_{B-L})$ 
for $T_{c}\sim 100\ {\rm GeV}$. This analysis explicitly demonstrates 
how a primordial $B-L$ is damped as the temperature falls from $T_{B-L}$ 
to $T_c$. In conclusion, the underlying brane model must accommodate
Yukawa couplings, which are either sufficiently large so that the reverse
proccess (restoring the lepton asymmetry above $T_c$) are operative or so
small so that both forward and reverse processes are frozen out in order
to prevent the primordial
$B-L$ from being washed out. The latter condition is clearly more
severe when the number of accessible KK modes is increased. This
conclusion is relevant only when the hierarchy among the  neutrino masses
is generated by the zero modes 
$f^{\ell\ \vec{0}}_{R}(\vec{y}_0)$.

\subsection{Hierarchy from the 5D Yukawa couplings}
A second possibility for having realistic
masses and mixings for neutrinos comes via the flavor structure
of the Yukawa matrix $\lambda^{\alpha \ell}$. Namely, while the
zero mode wave functions $f^{\ell\ \vec{0}}_{R}(\vec{y}_0)$ are 
of similar size, the entries of $\lambda^{\alpha \ell}$ may possess the
requisite  hierarchy in the 5D Yukawa couplings to explain the data.  
When the neutrino masses obey a hierarchical splitting,
$m_3\approx \delta m_{atm} \gg m_2 \approx \delta m_{sol} \gg m_1$,
so does the Yukawa matrix, $(h_{\vec{n}} h^{\dagger}_{\vec{n}})_{3 3} \gg
(h_{\vec{n}} h^{\dagger}_{\vec{n}})_{2 2} \gg (h_{\vec{n}} 
h^{\dagger}_{\vec{n}})_{11}$. 
In this case, we must consider separately the three forward and reverse
processes. For each generation the ratio of the two is still $N_{KK}$. 
Unless all three generations, are either completely in (or out) of
equilibrium, a prior asymmetry can be regenerated.   Consequently,
even if the heavier neutrinos can equilibrate with
the Higgs boson, $\mu_{\nu_{\mu}}=\mu_{\nu_{\tau}}=-\mu_{H}$,
the primordial asymmetry accumulated in $\nu_e$ survives the 
sphaleron reprocessing. Indeed, at temperatures
$T\sim T_c$, there remains a nonvanishing $B-L \propto  \mu_{\nu_e} +
\mu_H$ which sources the observed BAU.  This is precisely the scenario
developed in \cite{cdeo} in which an individual lepton flavor asymmetry
can be responsible for the BAU so long as one or two families are in (out) of
equilibrium {\em even if} the total $B-L$ = 0.

 In order to obtain
the analogous constraint to Eq. (\ref{rapidr}) due to the absence of any
effective lepton number violation one should replace
$h$ with $h_1$, i.e., the smallest Yukawa coupling.  In this case, all
three forward and reverse process will be faster than $H$.
Similarly, the analogous constraint to Eq. (\ref{rapidf}) due to freezing
out all of the interactions involving $\nu_R$ is obtained by replacing
$h$ with $h_3$. If we maintain that $h_1^2 \approx 10^{-5} h_3^2$, then
the bounds on $h_1$ can be read from Fig.~\ref{fig2} by shifting down
the upper boundary by five orders of magnitude.

\section{Specific Brane Models}
Independent of the mechanism for generating the hierarchy of the
neutrino masses, the survival of the primordial $B-L$ from erasure by
right-handed neutrino emission can only be determined only after
the KK mass spectrum
$m_{\vec{n}}$ and Yukawa spectrum $h_{\vec{n}}$ for the bulk leptons are
specified.  Thus in a given model in which the evolution of the
universe can be tracked up to temperatures above $T_c$, the
fate of the $B-L$ asymmetry can be determined from equation
(\ref{dense2}). The discussion in the last section is valid for any brane 
construction in which the SM spectrum is localized. It is therefore
convenient to discuss the phenomenon of $B-L$-violation
in specific brane models and show how $B-L$ evolves in the presence of
right-handed neutrino emission from the brane.

\subsection{Large Extra Dimensions}
In higher dimensional theories with flat and compact extra 
dimensions, a solution to the hierarchy problem requires
the extra dimensions to be large~\cite{gia1}, and therefore
they drastically change the early history of the Universe. In the 
absence of a mechanism which stabilizes the large extra 
dimensions, the evolution of the universe may not be standard much
above the nucleosynthesis era \cite{gia2}. Therefore, 
temperatures above the ${\rm MeV}$ scale may not be accessible, 
and the generation or erasure of the BAU cannot be discussed
in the context of these models. This is because unless the temperature of
the Universe  rises to some $T_{brane}$ above $T_c$, the processes we are
describing  do not occur. We note that the mechanism described in
\cite{pil}, in which leptogenesis occurs at temperatures much below $T_c$
overestimates the final baryon asymmetry by a very large factor. There
the asymmetry was estimated as being proportional to $e^{-T_c/T_{brane}}$,
and has neglected the factor of $\sim 8 \pi / g_W$ in the exponent
\cite{am}.  Alternatively, low scale leptogenesis is possible in orbifold
GUT  models~\cite{hmy}.

\subsection{Warped Extra Dimensions}
Another higher dimensional scenario which can solve the 
hierarchy problem is due to \cite{rs1} where two branes,
the Planck and visible (${\rm TeV}$) branes, are immersed in
the AdS$_{5}$ bulk at respective positions $y=0$ and $y=\pi r_c$.
The extra dimension $y$ is parametrized as 
$y=\pi \phi$ with $-\pi \leq \phi \leq \pi$, and the 
points $(x,\phi)$ and $(x,-\phi)$ are identified. 
Here $r_c$ is the radius of the $S^{1}/Z_2$ orbifold, and it 
determines the size of the extra dimension. For generating 
the hierarchy, $M_{Pl} e^{- \pi k r_c} \equiv {\rm TeV}$,  
one needs  $r_c k \sim 10$ where $k$ is the AdS curvature. 
Suppression of the higher order curvature effects in 
the gravity sector requires $k$ to be smaller than the 
fundamental scale of gravity, $M_5\approx (M_{Pl}^{2} k)^{1/3}$. 
Typically $k\simlt \eta M_5$ where we have introduced the 
parametrization $\eta$, satsifying $0\leq \eta < 1$.

Unlike large extra dimensions, the warped geometries
allow for reheat temperatures  $\sim {\cal{O}}({\rm TeV})$
which is the characteristic energy scale of the visible brane,
and we will see that the number of accessible KK states 
is $N_{KK}\sim \eta^{-3/2}$.
At very high temperatures, $T\gg {\cal{O}}({\rm TeV})$, the 
visible brane simply does not exist as it is
either shielded by the AdS horizon or pushed away 
from its true configuration. As the temperature drops
the visible brane with a hot SM spectrum emerges at $T_{brane}\sim 
{\cal{O}}({\rm TeV})$ \cite{nima}. In particular, for the RS1 set-up, 
standard evolution remains valid up to corrections of
${\cal{O}}(\rho^{2})$
( see e.g. \cite{cline}). More generally for warped extra dimensions we
will distinguish two  possibilities for the localization of SM fields.

\subsubsection{Matter on the TeV-brane}
Here, we place all SM fields on the TeV-brane, as was the case in the
original RS1 model, and introduce right-handed neutrinos in the
bulk~\cite{grossman}.  A bulk fermion
$\Psi^{\ell}(x,y)$, with Dirac  mass term 
$\zeta_{\ell}\ k\ \mbox{sgn}(\phi)\ \overline{\Psi}^{\ell} \Psi^{\ell}$, 
is a  Dirac spinor obeying the chiral decomposition in (\ref{psi}).
The orbifold symmetry requires $f_{L}^{\ell\ n}(\phi)$ to be odd (even) and 
$f_R^{\ell\ n}(\phi)$ even (odd) under the $Z_2$ symmetry. Therefore, to
generate neutrino masses on the ${\rm TeV}$ brane, the appropriate
boundary  conditions are 
\begin{eqnarray}
f_{L}^{\ell\ n} (0) = f_{L}^{\ell\ n} (\pi) = 0~,
\end{eqnarray}
such that  $f_{R}^{\ell\ 0}(\phi)$ is localized on the Planck brane with a tiny tail 
on the visible brane \cite{grossman}. The Yukawa couplings are given by 
\begin{eqnarray}
\label{yukawa1}
f_{R}^{\ell\ 0}(\pi)= \sqrt{2 \zeta_{\ell} -1}\ e^{-(\zeta_{\ell}-1/2) 
\pi k r_c}\:\:\:,\:\:\: f_{R}^{\ell\ n\neq 0}(\pi) = \sqrt{2}~,
\end{eqnarray}
which indicates that the couplings of the zero modes are 
at least fifteen orders of magnitude smaller than those of the KK 
levels provided that $\zeta_{\ell}>1/2$. Namely,
the bulk fermions $\Psi^{\ell}(x,y)$ must have a mass parameter
larger than half the AdS curvature scale in order to obtain small numbers 
to generate neutrino masses. On the other hand, the masses of the 
KK levels obey
\begin{eqnarray}
\label{masses1}
m_{\ell\ n} =  x_{\ell\ n} k e^{- \pi k r_c} \;~,
\end{eqnarray}
where $J_{\zeta_{\ell}-1/2}(x_{\ell\ n})=0$ \cite{grossman}. For arbitrary 
$\eta$, one finds that 
$m_{\ell\ n}\approx x_{\ell\ n} \eta^{3/2}\ {\rm TeV}$ which gives 
$m_{1}\sim M_{Z}$ when $\eta\approx 0.1$. In general, the smaller the value
of $k$, the smaller $m_{\ell\ n}$ becomes and the number of KK states
excited at TeV temperatures increases; however, it is convenient to 
choose $\eta\sim 0.1$ as a moderate value in order to suppress the AdS 
curvature with respect to $M_5$.
In order for the coupling of the zero modes (\ref{yukawa}) to
approximate the existing oscillation data, the mixing with the higher
modes (\ref{angle}) must be small, and this can be satisfied if the
entries of $\lambda^{\alpha\ \ell}\lsim 10^{-2}$, or in other words, 
$\left(h_{n} h^{\dagger}_{n}\right)_{\alpha \alpha} \simlt 10^{-4}$.

Returning to the analysis of generic brane models in Sec.~2, we can now
discuss the possible $B-L$ erasure in the RS1 model. When the neutrino 
textures are generated by $f_{R}^{\ell\ 0}$ in (\ref{yukawa1}), we see from 
Fig.~\ref{fig2} that the entire $B-L$ asymmetry is preserved for 
$\eta\sim 0.1$. This is because 
the experimental bound \cite{paul} for suppressing 
$\tan\theta_{\alpha}$ in (\ref{angle}), $\left(h_{n}
h^{\dagger}_{n}\right)_{\alpha \alpha} \sim 10^{-4}$,   is roughly
eight orders of magnitude higher than the bound obtained from 
Fig.~\ref{fig2}, for $N_{KK}\sim \eta^{-3/2} \simeq 30$.

On the other hand, as mentioned in Sec.~2.2, 
when all $f_{R}^{\ell\ n}$ are similar in size but the entries
of $\lambda^{\alpha \ell}$ possess a hierarchy to generate
neutrino masses, it is possible to suppress 
$\left(h_{n} h^{\dagger}_{n}\right)_{\alpha \alpha}$ for
the lightest flavor for all $n$. Then not only is $B-L$ preserved, even
if $(B-L)_{total} = 0$, a baryon asymmetry will be generated if the flavor
asymmetries do not all vanish \cite{cdeo} as would be the case for Yukawa
couplings in the small window for $N_{KK}\sim 30$.
{}From Fig.~\ref{fig2} we see that the window becomes larger and more than
one flavor symmetry could be washed out if $N_{KK}$ becomes very large.
Eventually for large enough $N_{KK}$, all three generations are washed away, 
and the final baryon asymmetry is erased.
 
In discussing the erasure of $B-L$ we have left unspecified 
the origin of the primordial asymmetry. The visible brane
has the natural scale ${\cal{O}}({\rm TeV})$ so
that it becomes more difficult to induce leptogenesis with
genuinely four-dimensional mechanisms with heavy Majorana neutrino
masses. However, since the sphaleron constraint only imposes a
bound on $h^2/M$, it is possible by tuning the Yukawa textures to
achieve leptogenesis with lighter, O(TeV), neutrino masses
\cite{luty}.

\subsubsection{Matter on the Planck-brane}
An alternative possibility is to assume that matter is localized
on the Planck-brane. In this scenario supersymmetry is now required to 
protect the Higgs mass, and supersymmetry is broken on the TeV-brane.
The warp factor is responsible for naturally explaining 
the TeV supersymmetry breaking scale~\cite{gp}.
In this set-up the appropriate boundary conditions are 
\begin{eqnarray}
f_{R}^{\ell\ n}(0)=f_{R}^{\ell\ n}(\pi)=0~,
\end{eqnarray}
where $f_L^{\ell\ 0}(\phi)$ is even under the orbifold symmetry and
$f_L^{\ell\ 0}(\phi=0)$ generates the Yukawa couplings (\ref{lag})
with $\psi_{R}^{\ell\ n}$ is replaced by $\psi_{L}^{\ell\ n\ c}$,
$c$ for charge conjugation.  
The KK mass spectrum has the same form as (\ref{masses1})
with  $x_{\ell\ n}$ being the roots of $J_{\zeta_{\ell}+1/2}(x)$ instead.
For $\zeta_{\ell}\geq -1/2$ the brane and bulk leptons are weakly coupled 
\begin{eqnarray}
\label{yukawa2}
f_{L}^{\ell\ 0}(0) =  \sqrt{1+2 \zeta_{\ell}} e^{- (\zeta_{\ell} + 1/2) 
\pi k r_c} \:\:\:,\:\:\:
f_{L}^{\ell\ n\neq 0}(0) = \frac{\sqrt{2}}{\Gamma(\zeta_{\ell}+1/2)}\
\left(\frac{x_{\ell\ n}}{2}\right)^{\zeta_{\ell}-1/2}\
\frac{e^{- (\zeta_{\ell} + 1/2) \pi k r_c}}{J_{\zeta-1/2}(x_{\ell\ n})}~,
\end{eqnarray}
for both the zero modes and higher KK states. 
Suppose that at least the first KK level 
is sufficiently light: $m_1\sim M_Z$ as was computed in the RS1
case. The smallness of the zero modes couplings in (\ref{yukawa2})
can be used to obtain a good fit to the neutrino oscillation data \cite{paul}
with small mixing with the higher KK levels. Then,
assuming that the standard FRW evolution holds up to temperatures
$T_{B-L}\sim {\rm TeV}$, one observes that the primordial $B-L$ asymmetry
is  not washed out. This is clear from (\ref{dense2}) where 
\begin{eqnarray}
\sum_{\vec{n}} \left(h_{\vec{n}} h^{\dagger}_{\vec{n}}\right)_{\alpha \alpha}
\sim \lambda^{\alpha\ \ell} e^{-(2 \zeta_{\ell}+1) \pi k r_c}
{\lambda^{\dagger}}^{\ell\ \alpha}~,
\end{eqnarray}
is much smaller than $10^{-12}$, and so even the small window at 
$N_{KK}\sim 30$ for Yukawa couplings $h^2\sim 10^{-12}$ is not a problem.
Therefore, by localizing matter on the Planck-brane, and keeping the 
TeV-brane at a finite distance, it is possible to naturally 
generate neutrino masses and mixings with no erasure of the primordial 
asymmetry.

When $r_c\rightarrow \infty$, one recovers the original RS2 limit, in
which case the  cosmological evolution depends on a new scale $M_{c}\equiv
\sqrt{M_{Pl} k}$, and it is only for temperatures $T\simlt M_{c}$
that the standard FRW cosmology is recovered. For higher temperatures,
the evolution does not have a FRW form and
${\cal{O}}(\rho^{2})$ contributions \cite{hebecker} are important. For
$M_c=10^{12}\ {\rm GeV}$, one has $M_{5}=10^{14}\ {\rm GeV}$ 
and $k=10^{6}\ {\rm GeV}$. On the other hand, $M_c$ can be lowered down 
to $1\ {\rm TeV}$ level for which $M_{5}=10^{8}\ {\rm GeV}$ 
and $k\sim m_{\nu}$ \cite{smallk}.

The form of the Yukawa couplings (\ref{yukawa2}) shows that 
the Dirac mass of the corresponding neutrino vanishes for 
infinitely separated branes. Hence the massive left-handed
neutrinos as well as their mixings cannot follow from brane-bulk
couplings in RS2 scheme. However, the vanishing of the Yukawa couplings as 
$r_c\rightarrow \infty$ does not mean that the $B-L$ damping 
rate (\ref{dense2}) vanishes. Indeed, at large $r_c$, many KK modes become 
kinematically accessible and they enhance the emission rate cumulatively 
as is the case for gravitons~\cite{hebecker,smallk}. In fact, using 
(\ref{yukawa2}) one obtains
\begin{eqnarray}
\sum_{\vec{n}} \left(h_{\vec{n}} h^{\dagger}_{\vec{n}}\right)_{\alpha \alpha}
\sim \lambda^{\alpha\ \ell} \left(\frac{M_c}{ 2 k}\right)^{2\zeta_{\ell}+1}
{\lambda^{\dagger}}^{\ell\ \alpha}~,
\end{eqnarray}
at $T\sim M_c$ on the brane. When $M_c$ is
at the intermediate scale, $M_c\sim 10^{12}\ {\rm GeV}$, this quantity
is approximately $10^{6 (2 \zeta_{\ell} +1)}$, which implies that, 
for $\zeta_{\ell}\sim {\cal{O}}(1)$, $\lambda^{\alpha \ \ell}\simlt 10^{-15}$
for the primordial $B-L$ in ${\bf L}^{\alpha}$ to remain in tact. 
However, when $\zeta_{\ell}\rightarrow -1/2$, 
the cumulative effect is mild, and it suffices to choose $\lambda^{\alpha\ 
\ell} \simlt 10^{-6}$ in order to preserve the primordial $B-L$. Thus,
even though bulk fermions are not responsible for generating the
masses and mixings of  the massive neutrinos, their emission from the
brane leads to a wash-out of  the primordial asymmetry.

The fact that energy scales much greater than ${\cal{O}} ({\rm TeV})$ 
are accessible on the RS2 brane enables one to invoke genuine 
four-dimensional mechanisms to generate neutrino masses and mixings as 
well as the lepton asymmetry in a natural way. Indeed, conventional Majorana 
leptogenesis in which the SM spectrum is augmented by three
brane-localized, right-handed neutrinos ${\bf N}^{\alpha}$
can be used \cite{fy}. If the ${\bf N}^{\alpha}$
have masses at the intermediate scale $M_{N}\sim M_c \sim 10^{10}\ {\rm
GeV}$ then $B-L$ is not erased and a finite BAU is generated so long as
the emission of right-handed neutrinos from the brane is suppressed.
Furthermore we note that in compact hyperbolic extra dimensions
where standard FRW evolution continues up to high temperatures, the weak
scale KK spectrum is similar to that in RS1, and any primordial 
$B-L$ will be erased unless the brane-bulk couplings are tuned 
accordingly~\cite{hyperbolic}.

\section{Conclusion}

This letter has been devoted to a discussion of $B-L$ violation induced
by the emission of right-handed neutrinos from the brane. The main
statement  of the analysis is that unless the $\nu_R$ production and its
reverse process are {\em both} fast or slow compared with the expansion
rate at $T_c \sim 100$ GeV, the
$B-L$ asymmetry of the brane can be washed out by the  emission of
right-handed neutrinos. A symmetry principle which can halt  the neutrino
emission from the brane could come through promotion of the global $B-L$
to a local invariance. However, this symmetry must be broken at  an
intermediate scale, and therefore, it may not be generically realizable
in  general brane models.

Among the brane world examples, it is not possible to discuss
the preservation of a primordial asymmetry in the large extra dimensions
since the temperature cannot be much above the ${\rm MeV}$ scale. 
In the RS geometry with 
matter localized on the TeV-brane, the bulk leptons are responsible for
generating the neutrino masses and mixings. If the bulk field
wavefunctions  generate the neutrino mass hierarchy, then $B-L$ erasure
only occurs for very large $N_{KK}$. However, when the neutrino mass
splittings  come from the brane-bulk couplings, then the lepton
asymmetry in the lightest (or massless) flavor can remain preserved.
Instead, if matter is localized on the Planck brane, we have 
the unique feature of generating both the neutrino masses and the
primordial $B-L$ via the usual leptogenesis mechanism. The $B-L$ 
asymmetry generated this way is not washed out via the emission of the 
bulk leptons from the brane because the brane-bulk couplings can 
naturally be made exceedingly small.

\section*{Acknowledgements}
We thank Gia Dvali, Gregory Gabadadze, Kimmo Kainulainen, Maxim
Pospelov, and Misha Voloshin for  useful discussions, and Jim Cline
and Elias Kiritsis for useful e-mail exchange.  We especially thank Daniel
Chung for many fruitful discussions and e-mail exchange.
This work was supported in part by 
DOE grant DE-FG02-94ER40823 at the University of Minnesota.

\end{document}